\documentclass[10pt, titlepage]{article}
\usepackage{epsf,here,amssymb}
\def\commentout#1{}
\newtheorem{fact}{Fact}

\def\proof{\noindent{\bf Proof.\ }}
\def\endproof{\hfill $\Box$}
\input epsf
\newtheorem{theorem}{Theorem}
\newtheorem{lemma}[theorem]{Lemma}
\newtheorem{corollary}[theorem]{Corollary}

\begin{document}
\begin{titlepage}
\title{\Large Probabilistic Analysis of Rule 2 }
\author{Jennie C. Hansen\\
\small Actuarial Math and Statistics Department\\
\small Herriot-Watt University\\
\small J.Hansen@ma.hw.ac.uk
 \and
Eric Schmutz\\
\small Department of Mathematics\\
\small Drexel University\\
\small Philadelphia, Pa. 19104 \\
\small Eric.Jonathan.Schmutz@drexel.edu\\
\\
Li Sheng\\
 \small  Department of Mathematics\\
 \small Drexel University\\
 \small Philadelphia, Pa. 19104\\
 \small lsheng@cs.drexel.edu\\
 }

\date{\today}
\maketitle \abstract{ Li and Wu proposed Rule 2,
 a  localized
 approximation
algorithm that attempts to find a small connected dominating set
in a graph. Here we study the asymptotic performance of Rule $2$
on random unit disk graphs formed from $n$ random points in an
$\ell_{n}\times \ell_{n}$ square region of the plane. If
$\ell_{n}=O(\sqrt{n/\log n})$, Rule 2 produces a dominating set
whose expected size is $O(n/(\log\log n)^{3/2}).$

\noindent {\bf keywords and phrases}: {\em  coverage process,
dominating set, localized algorithm, performance analysis,
probabilistic analysis, Rule k, unit disk graph} }
\end{titlepage}

\section{ Introduction}
Suppose random points $V_{1},V_{2},\dots ,V_{n}$ are selected from
a connected region ${\cal Q}$ in $\Re^{2}.$  For each $i$, let
$D_{1}(V_{i})$ be the unit disk centered at $V_i$. There is a
large literature on coverage processes\cite{Hall} that enables one
to answer questions such as whether or not the random disks are
likely to cover all of ${\cal Q}$, i.e. whether ${\cal Q}\subseteq
\bigcup\limits_{i=1}^{n}D_{1}(V_{i}).$ A variant question asks
whether there is small {\em subset} of the disks whose union
already covers ${\cal Q}$: given $k<n$, are there indices
$i_1<i_2<\dots i_{k}$ such that  ${\cal Q}\subseteq
\bigcup\limits_{j=1}^{k}D_{1}(V_{i_{j}}).$
 For this variant, there are several interesting ways to
modify the meaning of \lq\lq coverage.\rq\rq For example:  is
there a small subset of the disks whose union is connected and
contains all $n$ points $V_1,V_2,\dots ,V_n$ (but not necessarily
all of ${\cal Q}$)? These questions are a bit vague, but specific
examples arise naturally in connection with probabilistic models
for wireless networks. In particular, they are central to the
probabilistic analysis of Rule 2 in this paper.

  Rule 2 is a well known algorithm that was proposed by Wu and Li
\cite{WuLi} as a means of increasing the efficiency of routing in
ad hoc wireless networks. To describe the algorithm and a
probabilistic model, we need some graph theoretic terminology.  A
{\em unit disk graph } has for its vertex set ${\cal V}$ a finite
set of points in $\Re^{2}.$ Given the vertex set ${\cal V}$, the
edge set ${\cal E}$ is determined as follows: an {\it undirected}
edge $e\in {\cal E}$ connects vertices $u,v\in {\cal V}$  (and in
this case we say that $u$ and $v$ are adjacent) iff  $d(u,v)$, the
Euclidean distance between them, is less than one. Unit disk
graphs have been used by many authors
 as  mathematical
models for the interconnections between nodes in a wireless
network, and random unit disk graphs have been used as
probabilistic  models for these networks
\cite{CCJ},\cite{Gilbert},
\cite{GuptaKumar},\cite{Hale},\cite{HS},\cite{M},\cite{M2}. A {\sl
dominating set} in any graph $G=({\cal V},{\cal E})$ is a subset
${\cal C}\subseteq {\cal V}$  such that every vertex $v\in {\cal
V}$ either is in the set ${\cal C}$, or is adjacent to a vertex in
${\cal C}.$ We say ${\cal C}$ is a {\em connected dominating set }
if ${\cal C}$ is a dominating set and the subgraph induced by
${\cal C}$ is connected. Of course it is not possible for $G$ to
have a connected dominating set if $G$ itself is not connected. We
use the acronym \lq\lq CDS\rq\rq for a dominating set ${\cal C}$
such that the subgraph induced by ${\cal C}$ has the same number
of components that $G$ has. This paper deals with a {\em random}
unit disk graph model, ${\cal G}_{n}$, which is connected with
asymptotic probability one. Thus any CDS for ${\cal G}_{n}$ will
also be connected with high probability. We assume that each
vertex has a unique identifier taken from a totally ordered set.
For convenience, when $|{\cal V}|=n$, we will use the numbers
$1,2,\dots ,n$ as IDs, and will number the vertices accordingly.
 If $v_i$ is any vertex (with ID $i$), define the neighborhood
  ${\cal N}(v_i)$ to be
 the set consisting of $v_i$ and any vertices in ${\cal V}$ that are
 adjacent to $v_{i}.$
The CDS constructed by the Rule $2$ algorithm is denoted ${\cal
C}({\cal V})$, and its cardinality is $C({\cal V})=|{\cal C}({\cal
V})|.$ The elements of ${\cal C}({\cal V})$ are called \lq\lq
gateway nodes\rq\rq. ${\cal C}({\cal V})$ consists of all vertices
$v_{i}\in {\cal V} $ that are not excluded under the following
version of Rule 2:

 \vskip.5cm \noindent {\bf Rule 2: } {\sl Vertex
$v_{i}$ is excluded from ${\cal C}({\cal V})$  iff
 ${\cal N}(v_i)$ contains at least one set of two vertices
$v_{i_1},v_{i_2}$ such that
\begin{itemize}
\item    $i_{1}>i_{2}>i $ and
\item ${\cal N}(v_i)\subseteq {\cal N}(v_{i_1})\cup {\cal N}(v_{i_2})$ and
\item  $v_{i_{1}}$ is adjacent to $ v_{i_{2}}.$
\end{itemize}}

Wu and Li showed that this algorithm produces a CDS. They also
conjectured, based on simulation data, that it is  effective in
the sense that it selects a CDS that is small relative
to $n$ \lq\lq in the average
case\rq\rq. In this paper we treat the analysis of Rule 2
mathematicially by considering its performance when
it is applied to a random unit disk
graph ${\cal G}_{n}.$
 Specifically, let $\ell_{1}\leq
\ell_{2}\leq \dots $ be a sequence of real numbers such that
$\ell_{n}=O(\sqrt{n/\log n})$ as $n\rightarrow\infty,$ but
$\ell_{n}\geq \log n$ for all $n$. Let ${\cal Q}_{n}$ be an
$\ell_n\times \ell_{n}$ square region in $\Re^{2}$.
 Select
$n$ points $V_1,V_2,\dots  ,V_n$ independently and uniform
randomly from an ${\cal Q}_{n}$, and use these $n$ points as the
vertex set for a unit disk graph ${\cal G}_{n}$.  With this
probabilistic model, the  size of the Rule 2 dominating set is a
random variable. We prove asymptotic estimates for
the expected size of the Rule 2 dominating set.  The proof
involves some interesting problems in elementary geometry and
geometric probability.

\section{A Geometric Lemma}
As observed in
  \cite{JLMV},  a unit disk centered at a point $o$ cannot be completely covered with
  two unit  disks having centers at points $u$ and $w$
  ($u\not=o\not=w$): \ $(D_{1}(u)\bigcup D_{1}(w))^{c}\bigcap
  D_{1}(o)\not= \emptyset.$
   One might  infer that a typical vertex $o$ is not likely to be  be
   pruned under Rule 2
   because no two points in  ${\cal N}(o)$ will cover all the vertices in ${\cal N}(o).$
   This reasoning suggests that  Rule 2 will be ineffective.
   But such reasoning is not sound. Typically there are points $u$ and $w$
  that cover all but a negligible fraction of
  the disk centered at $o$. The uncovered  region is small enough
  so that it usually does not include any nodes.
A more precise version of this statement is proved in  the next
section, but first we need to look carefully at the area of regions
such as $(D_{1}(u)\bigcup D_{1}(w))^{c}\bigcap
  D_{1}(o).$ In particular, we  need Lemma
  \ref{notrunc}, which is the main result in this section.

 To state Lemma \ref{notrunc} we adopt some notation. Throughout
this section $b>1$ will be a parameter and in terms of $b$
we let  $L=\lfloor b^{1/3}(\log b)^{2}\rfloor, \delta={1\over
\sqrt[3]{b}\log b}$,   and  $\theta_{b}={\pi/L}.$  We fix
$o=(x_{o},y_{o})\in\Re^{2}$ 
and for any $r>0$,  
let $D_{r}(o)$ be the
closed disk centered at $o$ with radius $r$.
We
are going to partition the small disk $D_{\delta}(o)$  into $2L$
sectors as follows. Choose a new coordinate system centered at $o$, and for
$0\leq i< L,$ let
 $Q_{i}$ be the sector consisting  of those points
$(x,y)=(r\cos\theta,r\sin\theta)$ whose polar coordinates satisfy
$0<r\leq \delta $ and $(i-{1\over 2})\theta_{b}\leq \theta\leq
(i+{1\over 2}) \theta_{b} $.
 Similarly let $R_{i}$ be the sector
that is obtained by reflecting $Q_{i}$ about $o$, namely the
points with $0<r<\delta $ and $(i-{1\over 2})\theta_{b}<
\theta-\pi < (i+{1\over 2})\theta_{b} $.
 The analysis of Rule 2 depends on a geometric lemma about these sectors.
 For any $i$, and any points
$q_i\in Q_{i},u_i\in R_{i}$, let
$X(q_i,u_i)$ be the area of $(D_{1}(q_i)\bigcup
D_{1}(u_i))^{c}\bigcap D_{1}(o)$, i.e. the area of the omitted
region in $D_{1}(o)$ that is not covered by $(D_{1}(q_i)\bigcup
D_{1}(u_i)$ . Let $\tilde{q}_i$
 and
 $\tilde{u}_i$ be the extreme points whose polar coordinates are
 respectively
 $(r,\theta)=(\delta,(i-{1\over
2})\theta_{b})$ and $(r,\theta)=(\delta,(i+{1\over
2})\theta_{b}+\pi).$
We prove:

\begin{lemma}
\label{notrunc}
 There is a uniform constant $C>0$ such that, for $0\leq
i<L$, and for all $q_i\in Q_{i},u_i\in R_{i},$ we have
$
{X}(q_{i},u_{i})\leq X(\tilde{q}_{i},\tilde{u}_{i}) \leq
{C\over b\log^{3}b}$.
\end{lemma}

\begin{proof}

We prove four facts which together imply Lemma \ref{notrunc}. In
the first fact, we observe that omitted area $X(q,u)$ gets larger
if we move one (or both) of the two points $q,u$ away from the
origin along a radial line.

\begin{fact}
 Let $q_1,q_2$ and $u_1,u_2$ be four points in $D_{1}(o)$ such
that $q_1$ lies on the line segment $\overline{o,q_{2}}$ and $u_1$
lies on the line segment $\overline{o,u_2}$. Then $X(q_2,u_2)\ge
X(q_1,u_1)$.
\end{fact}

\proof It suffices to show that  $D_{1}(q_{2})\cap D_{1}(o)
\subseteq D_{1}(q_1)\cap D_{1}(o)$ and  that $D_{1}(u_{2})\cap
D_{1}(o) \subseteq D_{1}(u_1)\cap D_{1}(o).$
 Suppose $p\in D_{1}(q_{2})\cap D_{1}(o).$ Since $q_1$
lies on the line segment from $o$ to $q_2$, we have $d(q_1,p)\leq
\max(d(o,p),d(q_2,p)) \leq 1.$
 Hence $p\in D_{1}(q_1)\cap D_{1}(o).$
 By a  similar same argument,
 $D_{1}(q_{2})\cap D_{1}(o) \subseteq D_{1}(q_1)\cap
D_{1}(o).$

\endproof

\begin{fact}\label{f2}
  Let $a,b$ be the two points where the circles $\partial
D_{1}(p),
\partial D_{1}(q)$ intersect. Then, $\overline{a,b}\ \bot \
\overline{p,q}$, and  the two line segments $\overline{a,b}$ and
$\overline{p,q}$  intersect at their midpoints.
\end{fact}

\proof This follows immediately from the fact that $d(p,a)=d(p,b)=
d(q,a)=d(q,b)=1$.
\endproof

\begin{fact}
Let $o_1,o_2$ be two points on the circle $x^2+y^2=\delta^2$.
Then, $X(o_1,o_2)$  is a decreasing function of $\angle o_1oo_2$.
\end{fact}

\proof For convenience, we will use polar coordinates. Without
loss of generality, let $o_1$ be the point with polar coordinates
$(r_{o_1},\phi_{o_1})=(\delta,\pi)$. Let $o_2$ be an arbitrary
point on the circle with the polar coordinates $(\delta,\phi_2)$.
By symmetry, we only need to consider the case when $o_2$ is in
the first or second quadrant; we may, without loss of generality,
assume that $0\le \phi_2 \le \pi$.  We will show that $X(o_1,o_2)$
is an increasing function of $\phi_2$, then the result follows
from the fact that $\angle o_1oo_2=\pi-\phi_2$.

Let $a_1,b_1$ be the two points where the circles $\partial
D_{1}(o_1)$ and $\partial D_{1}(o)$ intersect, with $a_1$ in the
second quadrant and $b_1$ in the third quadrant.

Let $o^*$ be a point on the circle $x^2+y^2=\delta^2$ so that
$\partial D_{1}(o^*)$ meets with both $\partial D_{1}(o)$ and
$\partial D_{1}(o_1)$ at $a_1$.  Let $b^*,d^*$ be the other
intersection points of $\partial D_{1}(o^*)$ with $\partial
D_{1}(o)$ and $\partial D_{1}(o_1)$, respectively. For
convenience, let's denote $\phi_{o^*}$ by $\phi^*$.
Figure~\ref{fig:phi} illustrates the position of $\partial
D_{1}(o_1),\partial D(o)$, and $\partial D_{1}(o^*)$ and their
intersections.

\begin{figure}[H]
\begin{center}
\leavevmode \epsfysize=2in \epsfbox{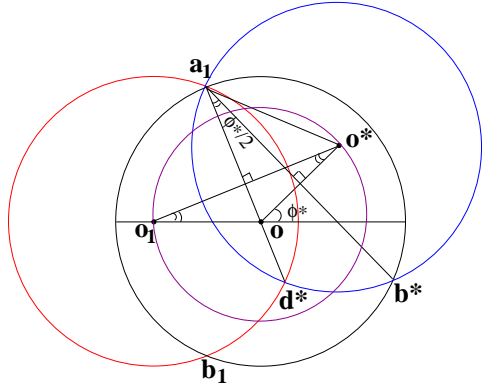}
\end{center}
\caption{The position of the circle $\partial D_{1}(o^*)$}
\label{fig:phi}
\end{figure}

As in the proof of  Fact~\ref{f2},we have  $\overline{a_1,d^*} \
\bot \ \overline{o_1,o^*}$, $\overline{a_1,b^*}\ \bot \
\overline{o,o^*}$. Notice also that $o$ is on the line segment
$\overline{a_1,d^*}$. So,
\begin{equation}~\label{e1}
\angle b^*a_1o=\angle oo^*o_1=\angle o^*o_1o=\frac{\phi^*}{2}.
\end{equation}
It follows that
\begin{equation}\label{sin-phi*}
0< \phi^*/2 <\pi /2,\ \mbox{and},\ \sin \frac{\phi^*}{2}=
\frac{\delta}{2}
\end{equation}
Now, for the point $o_2$ with polar coordinates $(\delta,\phi_2)$,
let $a_2,b_2$ denote the two points where $\partial D_{1}(o_2)$
and $\partial D_{1}(o)$ intersect, and let $c_2,d_2$ denote the
two points where $\partial D_{1}(o_2)$ and $\partial D_{1}(o_1)$
intersect. There are  two cases to consider: $\phi_2 \leq \phi^*$,
and  $\phi_2 \geq \phi^*$

Case 1. $\phi_2 \le \phi^*$.

\begin{figure}[H]
\begin{center}
\leavevmode \epsfysize=2in \epsfbox{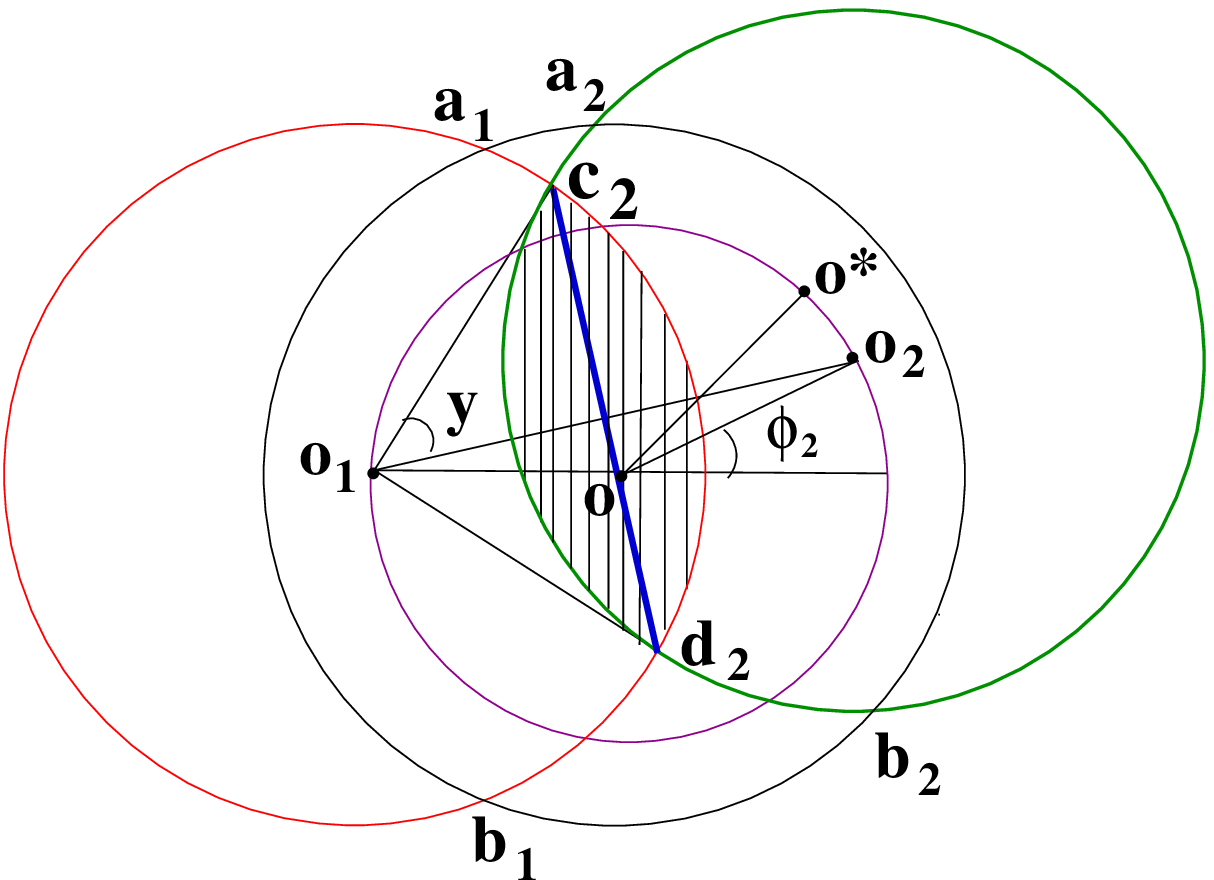}
\end{center}
\caption{The case when $\phi_2\le \phi^*$} \label{case1}
\end{figure}

 Notice that $a_1,b_1$ partitions the circle $\partial D_{1}(o)$
into two arcs: the right section and the left section. When,
$\phi_2 \le \phi^*$, as illustrated in Figure~\ref{case1}, $a_2,
b_2$ are both on the right section of the circle $\partial
D_{1}(o)$ between $a_1,b_1$. Similarly, $c_2,d_2$ are both  on the
right section of the circle $\partial D_{1}(o_1)$ between
$a_1,b_1$. Clearly,
\[
X(o_1,o_2)=B_1-(B_2-B_3)=B_1-B_2+B_3,\] where
\begin{itemize}
\item $B_1=area(D_{1}(o_1)^c\cap D_{1}(o))$
\item $B_2=area (D_{1}(o)\cap D_{1}(o_2))$
\item $B_3=area(D_{1}(o_1)\cap D_{1}(o_2))$, the shaded area in
  Figure~\ref{case1}
\end{itemize}
Notice that $B_3$ is the only area that depends on $\phi_2$. We
shall now give an expression for $B_3$.

Let's denote $\angle c_2o_1o_2=y$. Since $\angle
o_2o_1o=\frac{\phi_2}{2}$, we have
\begin{equation}~\label{ey}
0< y< \frac{\pi}{2},\ \mbox{and}, \ \cos y=\delta \cos
\frac{\phi_2}{2}
\end{equation}

By symmetry, one can see that the shaded region is partitioned
equally  by the line $\overline{c_2,d_2}$. So,
\[B_3= 2(\frac{2y}{2\pi} \pi -\frac{1}{2} (2\sin y)(\cos
y))=2y-\sin 2y. \] Here, the first term is the area of the sector
$D_{1}(o_1)$ that extends from $c_2$ to $d_2$, and the second term
is the area of the triangle($c_2,o_1,d_2$).

From the above two equations, we have
\[\frac{d X(o_1,o_2)}{d \phi_2}=\frac{d B_3}{d \phi_2}
=\frac{d B_3}{d y} \cdot \frac{d y}{d \phi_2}= (1-\cos 2y)\cdot
{\delta\sin{\phi_{2}\over 2}\over 2\sin y} > 0.\]
Here the last inequality follows from the fact that
$0<\frac{\phi_2}{2},y < \frac{\pi}{2}$. Thus
 $X(o_1,o_2)$ is an increasing function in $\phi_2$.

Case 2. $\phi_2 > \phi^*$.
\begin{figure}[H]
\begin{center}
\leavevmode \epsfysize=2in \epsfbox{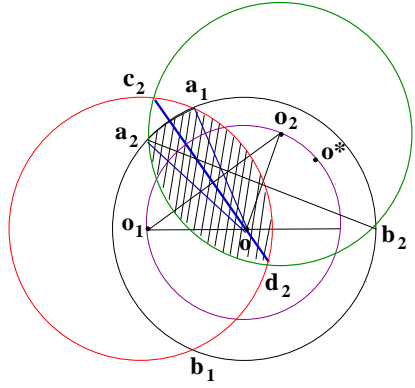}
\end{center}
\caption{The case when $\phi_2> \phi^*$} \label{case2}
\end{figure}

One can see from Figure~\ref{case2} that
\[
X(o_1,o_2)=B_1-(B_2-B_3)=B_1-B_2+B_3\] Where $B_1,B_2$ are defined
the same as those in the case 1, but
\[
B_3=area(D_{1}(o_1)\cap D_{1}(o_2)\cap D_{1}(o)), \ \mbox{the
shaded area in
  Figure~\ref{case2} }\]
Again, $B_3$ is the only area that depends on $\phi_2$. We will
now give an expression of $B_3$.

We show first that $\angle c_2oa_1 = \angle a_2oc_2$ by showing
that $\phi_{c_2}-\phi_{a_1}=\phi_{a_2}-\phi_{c_2}$. Then, it
follows that  $B_3$ is split in half by the line segment
$\overline{c_2,d_2}$.

>From Figure~\ref{fig:phi}, one can see that
\begin{equation}
\phi_{a_1}=\phi^*+(\frac{\pi}{2}-\angle
b^*a_1o)=\phi^*+(\frac{\pi}{2}-\frac{\phi^*}{2})=\frac{\pi}{2}+\frac{\phi^*}{2}
\end{equation}

To find $\phi_{a_2}$, observe that, as in the proof of
Fact~\ref{f2},we have $\overline{a_2,b_2}\ \bot \
\overline{o,o_2}$. So, $\sin \angle b_2a_2o =\frac{\delta}{2}$.
Comparing with (\ref{sin-phi*}), we see that $\sin \angle b_2a_2o
= \sin \frac{\phi^*}{2}$. This implies that $\angle b_2a_2o=
\frac{\phi^*}{2}$. Thus,
\begin{equation}
\phi_{a_2}=\phi_2+(\frac{\pi}{2}-\angle
b_2a_2o)=\phi_2+(\frac{\pi}{2}-\frac{\phi^*}{2})
\end{equation}

Now, for $c_2$, using the fact that $\overline{c_2,o} \ \bot\
\overline{o_1,o_2}$,
\begin{equation}
\phi_{c_2}=\pi-(\frac{\pi}{2}-\angle
o_2o_1o)=\pi-(\frac{\pi}{2}-\frac{\phi_2}{2})=\frac{\pi}{2}+\frac{\phi_2}{2}
\end{equation}

It follows that
$\phi_{c_2}-\phi_{a_1}=\phi_{a_2}-\phi_{c_2}=\frac{\phi_2}{2}-\frac{\phi^*}{2}$.
Now, using that the circle $\partial D_{1}(o_1)$ in the polar
system is
\[r=\sqrt{1-\delta^2\sin ^2 \phi} -\delta \cos \phi\]
and that
\begin{equation}
\phi_{d_2}=-(\pi-\phi_{c_2})=-(\frac{\pi}{2}-\frac{\phi_2}{2})
\end{equation}
we get
\[\begin{array}{ll}
B_3&=2(
\int_{-(\frac{\pi}{2}-\frac{\phi_2}{2})}^{\frac{\pi}{2}+\frac{\phi^*}{2}}\int_{0}^{\sqrt{1-\delta^2\sin
    ^2 \phi} -\delta \cos \phi}
    r\,drd\phi+\frac{\frac{\phi_2}{2}-\frac{\phi^*}{2}}{2\pi}\cdot
    \pi
) \vspace{0.2cm}\\
 &=
 \int_{-(\frac{\pi}{2}-\frac{\phi_2}{2})}^{\frac{\pi}{2}+\frac{\phi^*}{2}}
1-\delta^2\sin^2 \phi+\delta^2\cos^2 \phi -2\delta \cos \phi
\sqrt{1-\delta^2\sin^2 \phi}
    d\phi  + \frac{{\phi_2}-{\phi^*}}{2}

\end{array}
\]

Thus,
\[\begin{array}{lll}
\frac{d X(o_1,o_2)}{d
  \phi_2}=\frac{d B_3}{d
  \phi_2}&=&-\frac{1}{2}[1-\delta^2\sin^2
  (-\frac{\pi}{2}+\frac{\phi_2}{2})+\delta^2\cos^2
  (-\frac{\pi}{2}+\frac{\phi_2}{2}) \vspace{0.2cm} \\
&& -2\delta\cos
  (-\frac{\pi}{2}+\frac{\phi_2}{2})\sqrt{1-\delta^2\sin^2
  (-\frac{\pi}{2}+\frac{\phi_2}{2})}] +\frac{1}{2}
\vspace{0.2cm} \\
 &=&\frac{1}{2}[\delta^2\cos^2\frac{\phi_2}{2}-\delta^2\sin^2\frac{\phi_2}{2}+
2\delta \sin \frac{\phi_2}{2}
  \sqrt{1-\delta^2\cos^2\frac{\phi_2}{2}}]\vspace{0.2cm} \\
&=&\frac{1}{2}[-(\delta\sin\frac{\phi_2}{2}
  -\sqrt{1-\delta^2\cos^2\frac{\phi_2}{2}})^2+1] \vspace{0.2cm} \\
&\ge &0
\end{array}
\]
The last inequality follows because $0\le \delta
\sin\frac{\phi_2}{2}\le 1$, $0\le
\sqrt{1-\delta^2\cos^2\frac{\phi_2}{2}}\le 1$, and thus
$(\delta\sin\frac{\phi_2}{2}
  -\sqrt{1-\delta^2\cos^2\frac{\phi_2}{2}})^2 <1$.

\end{proof}
\begin{fact}
\label{omitted}Uniformly for all $i$, we have
 $X(\tilde{q}_i,\tilde{u}_i)=O({1\over
b\log^{3}b}).$

\end{fact}
\begin{proof}
Without loss of generality, let $i=0$ and $v=(0,0).$ To simplify
notation, define $x_{b}=\delta\cos(-{1\over 2}\theta_{b})$,
$y_{b}=\delta\sin(-{1\over 2}\theta_{b}).$
 Let $(\xi,\eta)$ be the point in the first quadrant where
the circles $x^2+y^2=1$ and $(x-x_{b})^{2}+(y-y_{b})^2=1$ meet.
Then

$$X(\tilde{q}_0,\tilde{u}_0)\leq 4\int\limits_{0}^{\xi}
\sqrt{1-x^{2}}-(y_{b}+\sqrt{1-(x-x_{b})^{2}})dx$$
$$= -4y_{b}\xi+4\int\limits_{0}^{\xi}{
{-2xx_{b}+x_{b}^{2}}\over
\sqrt{1-x^{2}}+\sqrt{1-(x-x_{b})^{2}}}dx$$ Hence we have
\begin{equation}
\label{xbound} X(\tilde{q}_0,\tilde{u}_0)=O(\xi
y_{b})+O(x_{b}\xi^{2})+O(x_{b}^{2}\xi).
\end{equation}
Note that $x_{b}^{2}+y_{b}^{2}=\delta^{2}={1\over
b^{2/3}\log^{2}b},$ that $\xi^{2}+\eta^{2}=1$, that
$(\xi-x_{b})^{2}+(\eta-y_{b})^{2}=1$, that
 $x_{b}= \delta(1+O(\theta_{b}^{2}))$, and that $y_{b}=
{-\delta\theta_{b}\over 2}(1+O(\theta_{b}^{2})).$ Combining these
equations, we get
 $\xi =O(\delta)$.
  Putting this estimate back into (\ref{xbound}), we get

\begin{equation}
\label{xarea} X(\tilde{q}_0,\tilde{u}_0)=O({1\over b\log^{3}b}).
\end{equation}
\end{proof}

In the analysis of Rule 2 it is necessary to consider
vertices in ${\cal G}_n$ which are close to the boundary
of the square ${\cal Q}_n$. For this reason we
define, for $o\in \Re_+^2$, the ``truncated unit disk''
$\hat{D}_1(o):=D_1(o)\cap\Re_+^2$ and we note that
$\hat{D}_{1}(o) \subseteq D_{1}(o) ,$ and $\hat{D}_{1}(o)=
 D_{1}(o)$ iff $x_{o},y_{o}\geq 1.$
Then for $L$ and $\delta$ as defined above,
we have the following corollary to Lemma \ref{notrunc}:

\begin{corollary}
\label{trunc}
 There is a uniform constant $C>0$ such that,
for all $o\in\Re_+^2$ such that $D_{\delta}(o)\subseteq
\Re_+^2$, for $0\leq
i<L$, and for all $q_i\in Q_{i},u_i\in R_{i},$ we have
$\hat
{X}(q_{i},u_{i})\leq X(\tilde{q}_{i},\tilde{u}_{i}) \leq
{C\over b\log^{3}b}$, where $\hat{X}(q,u)$ is the area of
$(D_{1}(q)\cap D_{1}(u))^{c}\cap \hat{D}_{1}(o).$
\end{corollary}

\begin{proof}
Clearly $\hat{X}(q_i,u_i)\leq X({q}_{i},{u}_{i})$ since
$\hat{D}_{1}(o) \subseteq D_{1}(o)$.
So the result follows from Lemma  \ref{notrunc} (since
$\tilde{q}_i, \tilde{u}_i\in D_{\delta}(o)
\subseteq \Re_+^2$).
\end{proof}

\section{Local Coverage by Two Discs}

Recall that under Rule 2 a vertex $v_i$ is excluded from ${\cal C}({\cal V})$
if there are two adjacent vertices, $v_{i_1}, v_{i_2}\in {\cal N}(v_i)$,
 with higher IDs than
$v_i$ which also  `cover' $v_i$, i.e. ${\cal N}(v_i)\subseteq {\cal N}(v_{i_1})
\cup {\cal N}( v_{i_2})$). In the analysis of Rule 2 we will distinquish vertices
in ${\cal N}(v_i)$ with higher ID than $v_i$ by coloring them blue; all other
vertices in ${\cal N}(v_i)$ are colored white. With this in mind, we consider in this
section a two-colored random unit disk graph and prove a local coverage
result.

Let $w$ and $b$ be positive integers such that $w<b(\log b)^{2}$
and, as before, let $L=\lfloor b^{1/3}(\log b)^{3/2}\rfloor$ and $\delta=
{1\over b^{1/3}\log b}.$
 Fix $o\in\Re^{2}_{+}$ such that $D_{\delta}(o)\subseteq
\Re_+^2$ and select $w+b$ points
independently and uniform randomly from
the truncated disk $\hat{D}_{1}(o).$ Color
the first $w$ points white, and the remaining $b$ points blue.
 Form a random (improperly colored) unit disk graph $\hat{{\cal
H}}_{w,b}$ by putting an edge between two of the $w+b$ colored
points iff the distance between them is one or less.
 Our goal in this section is to prove that,
with high probability, $\hat{{\cal H}}_{w,b}$ contains a
 dominating set consisting of two blue vertices that are adjacent
 to each other.

For $0\leq i<L$, let $Q_i, R_i$ denote the sectors
of $D_{\delta}(o)$ as defined in the previous section
and let $N(Q_{i}), N(R_{i})$ respectively  be the number of blue
vertices of $\hat{{\cal H}}_{w,b}$ that lie in $Q_{i}$ and $R_{i}$.
 Let $\tau_{b}=\sum\limits_{i=0}^{L-1}I_{i}$ where, in this section only, the $I_{i}=1$
if and only if $N(R_{i})=N(Q_{i})=1$ (and  otherwise $I_i=0$.) We note
that the distribution of $\tau_b$ depends on the position of $o$
and we indicate this dependence by using the notation
${\Pr}_o(\tau_b\in \cdot)$. Provided $o$ is not too close to the boundary of $\Re_+^2$,
we can obtain uniform bounds on the tail of the distribution
of $\tau_b$:
\begin{lemma}
\label{taubound}
 ${\Pr}_o\left(\tau_b <{b^{1/3}\over 16\log^{6}b}\right) =
 O({\log ^{6}b\over b^{1/3}})$ uniformly for all $o\in \Re_+^2$
such that $D_{\delta}(o)\subseteq \Re_+^2$.
\end{lemma}
\begin{proof}
Let $|\hat{D}_{1}(o)|$ denote the area of $\hat{D}_{1}(o)$, let
$\hat{\lambda}=\hat{\lambda}(o)={\pi\over |\hat{D}_{1}(o)|},$
and define
\begin{equation}
\hat{p}={{\rm Area}(Q_{i})\over |\hat{D}_{1}(o)|} =
{\pi\delta^{2}/2L\over |\hat{D}_{1}(o)| }= {\hat{\lambda}\over
2b\log^{4}b}\left(1+O({1\over b^{1/3}\log^{2}b})\right).
\end{equation}
The expected value of $I_{i}$ depends on $o$:
\begin{equation} E_{o}(I_{i})=b(b-1)\hat{p}^{2}(1-2\hat{p})^{b-2}= {\hat{\lambda}^{2}\over
4\log^{8} b}\left(1+O({1\over \log^{4}b})\right).
\end{equation}
 Hence
\begin{equation}
\label{Etaub}
E_{o}(\tau_{b})=LE_{o}(I_{i})={b^{1/3}\hat{\lambda}^{2}\over
4\log^{6}b}\left(1+O({1\over \log^{4}b})\right).
\end{equation}
We likewise have, for $i\not=j$,
\begin{equation}
\label{crossterms}
E_{o}(I_{i}I_{j})=b(b-1)(b-2)(b-3)\hat{p}^{4}(1-4\hat{p})^{b-4}=
{\hat{\lambda}^{4}(o)\over 16\log^{16} b}\left(1+O({1\over
\log^{4}b})\right).
\end{equation}
Note that \begin{equation} \label{hatalphabnd}
 \pi \ge |\hat{D}_{1}(o)| \ge {\pi\over
4},\end{equation} and therefore \begin{equation}
\label{hatlambdabnd}
 1\leq
\hat{\lambda}(o)\leq 4.
\end{equation}
 Therefore we have uniformly for all $o\in\Re_+^2$ such that
$D_{\delta}(o)\subseteq \Re_+^2$
\begin{equation}\label{variance} Var(\tau_{b})=O\left({b^{1/3}\over \log^{6}b}
\right).
\end{equation}
Observe that
\begin{equation}
\label{chebyshev} {\Pr}_o\left(\tau_b <{b^{1/3}\over 16\log^{6}b}\right)\leq
{\Pr}_o\left(\tau_b \leq {1\over 2}E_o(\tau_{b})\right) \leq
{\Pr}_o\left(|\tau_{b}-E_o(\tau_{b})|>{1\over 2}E_o(\tau_{b})\right).
\end{equation}
The lemma now follows from (\ref{variance}), (\ref{chebyshev}) and
Chebyshev's inequality.
\end{proof}
\vskip.5cm Recall our assumptions that $w<b(\log b)^{3/2},$ that
$\delta={1\over b^{1/3}\log b},$ and that $x_{o},y_{o}\geq
\delta.$ With these assumptions, we have:

\begin{theorem}
\label{localCDS}
 There is a constant $c>0$, independent of the position of $o$, such that with probability at least $1-{c\over ( \log b)^{3/2}},$
 the random graph
$\hat{{\cal H}}_{w,b}$  has a connected dominating set that
consists of two blue vertices in $D_{\delta}(o).$
\end{theorem}

\begin{proof}


Let ${\cal T}_{b}\subseteq \bigl\lbrace 0,1,2,3,\dots
,L-1\bigr\rbrace $ be the random subset of indices such that $i\in
{\cal T}_{b}$ iff $N(Q_{i})=N(R_{i})=1.$  If
 ${\cal T}_{b}\not= \emptyset,$
define $Y=\min {\cal T}_{b}$ to be the smallest of the indices in
 ${\cal T}_{b}$; otherwise, if ${\cal T}_{b}=\emptyset,$ set $Y=-1$.

Define the random variable $X_b$ as follows: If $\tau_{b}=|{\cal
T}_{b}|=0 $ then $X_{b}=0$; otherwise, if
 ${\cal T}_{b}=\bigl\lbrace i_1,i_2,\dots
i_{\tau_{b}}\bigr\rbrace$ and $i_1<i_2<\dots <i_{\tau_{b}}, $
  then
$X_{b}=1$ iff $Q_{i_{1}}\cup R_{i_{1}}$ contains a blue connected
dominating set for $\hat{{\cal H}}_{w,b}.$

Let ${\cal B}=\bigl\lbrace g_{1},g_{2},\dots ,g_{b}\bigr\rbrace $
be the set of blue nodes, selected independently and uniform
randomly from $\hat{D}_{1}(o).$   Define ${\cal Z}={\cal
B}\bigcap D_{\delta}(o)$ to be set of blue points that fall near
the origin $o$, and let $Z=|{\cal Z}|$ be the number of these points.
Then

\begin{equation}
{\Pr}_o(X_{b}=0)\leq {\Pr}_o\left(X_{b}=0,\tau_{b}\not=0, Z\leq {2\hat{\lambda}
b^{1/3}\over (\log b)^{2}}\right)
+{\Pr}_o(\tau_{b}=0)+{\Pr}_o\left(Z>{2\hat{\lambda}b^{1/3}\over (\log b)^{2}}
\right).
\end{equation}
Note that $Z$ has a binomial distribution: $ Z{\buildrel d\over
=}Bin(b,\hat{\lambda}\delta^{2})$ where $\hat{\lambda}$ is as defined in the
proof of Lemma \ref{taubound}. If
$\beta={2\hat{\lambda}b^{1/3}\over (\log b)^{2}}$, then by
Chernoff's inequality,
\begin{equation}
\label{Binomial} {\Pr}_o(Z\geq \beta)
 \leq \exp(-b^{1/3}/4(\log b
)^{2}).
\end{equation}
By Lemma \ref{taubound}, ${\Pr}_o( \tau_{b}=0)= O({\log^{6}b\over
b^{1/3}}).$
 Therefore
\begin{equation}
\label{RHS}
\label{Xr } {\Pr}_o(X_{b}=0)\leq   {\Pr}_o(X_{b}=0,\tau_{b}\not=0,Z\leq
\beta) +O({\log^{6}b\over b^{1/3}}).
\end{equation}

 Now we decompose  the first term on the right side of (\ref{RHS}) according to
 the value of $Y$.
\begin{equation}
\label{onestar}
{\Pr}_o(X_{b}=0,\tau_{b}\not= 0, Z\leq \beta)=
\sum\limits_{k=0}^{L-1} {\Pr}_o( X_{b}=0| Y=k, Z\leq \beta){\Pr}_o(Y=k,
Z\leq \beta).
\end{equation}
(The redundant condition $\tau_{b}\not=0$ need not be included on
the right side of (\ref{onestar})   because it a consequence of
the condition $Y\geq 0.$) We have
\begin{equation}
{\Pr}_o( X_{b}=0| Y=k, Z\leq \beta) =\sum\limits_{S} {\Pr}_o(X_{b}=0|
{\cal Z}=S,Y=k) {\Pr}_o( {\cal Z}=S\bigl| Y=k, Z\leq \beta)
\end{equation}
where the sum is over subsets $S\subseteq [b]$ such that
$2\leq |S|\leq \beta .$

\begin{equation}
 \Pr(X_{b}=0| {\cal Z}=S,Y=k)=1-\Pr(X_{b}=1|{\cal Z}=S,Y=k),
\end{equation}
so it is enough to find a lower bound for $\Pr(X_{b}=1|{\cal
Z}=S,Y=k).$

\commentout{ By Boole's inequality
\begin{equation}
\label{Boole2}
 {\Pr}_o(X_{b}=0| {\cal Z}=S,{\cal Y}_{b}=Y)\leq
 \ell(1-{\Pr}_o(J_{i_{1}}=1|{\cal Z}=S,{\cal Y}_{b}=Y)).
\end{equation}
}

 To simplify notation, let $\gamma=X(\tilde{q}_0,\tilde{u}_0),$
and recall that $\gamma= O({1\over b\log^{3}b}).$
In this section of the paper, define $|{D}_{\delta}(o)|={\pi \over
b^{2/3}(\log b)^{2}}$ to be the area of the disk $D_{\delta}(o)$,
and let $|\hat{D}_{1}(o)|=Area(\hat{D}_{1}(o)).$ An important
observation is that, once we have specified  $b-|S|=$ the {
number} of blue points that fall {\em outside} $D_{\delta}(o)$,
the locations in $D_{\delta}(o)^{c}\cap\hat{D}_1(o)$  of these $b-|S|$ points are
independent of the locations of the $|S|$ blue points {\sl in}
$D_{\delta}(o),$ and are also independent of the locations of the
white points. Hence
 \begin{equation}
{\Pr}_o(X_{b}=1|{\cal Z}=S,Y=k)\geq {(1-{|{D}_{\delta}(o)|\over
|\hat{D}_{1}(o)|}-{\gamma\over |\hat{D}_{1}(o)| } )^{b-|S|} \over
(1-{|{D}_{\delta}(o)|\over |\hat{D}_{1}(o)|
})^{b-|S|}}\left(1-{\gamma\over |\hat{D}_{1}(o)| }\right)^{w}
\end{equation}
\begin{equation}
\geq \left(1-{C\over b(\log b)^{3}}\right)^{b-|S|+w}
\end{equation}
 for some constant $C$ that is independent of $o$. With our
assumption $w<b(\log b)^{3/2}$
we get, for all sufficiently large
$b$,  the lower bound
\begin{equation}
\label{oops}
 {\Pr}_o(X_{b}=1|{\cal Z}=S,Y=k) \geq
\left(1-{C^{'}\over b(\log b)^{3}}\right)^{b(\log b)^{3/2}} \geq 1-{C^{''}\over
(\log b)^{3/2}}
\end{equation}
for some constants $C'$ and $C''$ which are independent of ${\cal Z}, Y$, and $o$.
Hence
\begin{equation}
{\Pr}_o(X_{b}=0)\leq {c\over (\log
b)^{3/2}}
\end{equation}
for some constant $c$ that is {\em independent of the point $o$.}

\end{proof}

\section{Analysis of Rule 2}
Let ${U}$ be the number of nodes that become non-gateways when
Rule 2 is applied to the random graph ${\cal G}_{n}$: ${
U}=\sum\limits_{i}I_{i}$ where (in this section)
the indicator variable $I_i=1$ 
iff the node with ID $i$ becomes a non-gateway under
Rule $2$. Assume that there is a positive constant $\bar{c}$ such that,
for all $n>1$, $\log n \leq \ell_{n}\leq \bar{c}\sqrt{{n\over \log n}}$.
Let $\xi_{n}={\alpha_{n}\over \ell_{n}^{2}}$, where $\langle
\alpha_{n}\rangle$ is  any sequence of real numbers  satisfying
the following three conditions:
\begin{itemize}
 \item
$\alpha_{n}=o(n)$ as $n\rightarrow\infty.$
\item  $\xi_{n}={\alpha_{n}\over
\ell_{n}^{2}}\rightarrow\infty$ as $n\rightarrow\infty.$
\item For all sufficiently large $n$,
${16n\over \log^{3/2}\xi_{n}} < \alpha_{n}.$
\end{itemize}
 For example, if $\ell_{n}=\Theta(\sqrt{n/\log n}),$
 then the sequence
$\alpha_{n}={32n\over (\log\log n)^{3/2}}$
 satisfies the three
conditions. On the other hand, if $\ell_{n}=\Theta(({n/\log
n})^t)$ for some fixed positive $t<1/2$, then $\alpha_{n}={n\over
\log n}$ satisfies the three conditions above.  With these three
assumptions, our goal is to prove
\begin{theorem}
\label{main}
 $E({ U})\geq n -O(\alpha_{n}).$
\end{theorem}

\begin{proof}
 The idea of the proof is to use Theorem
\ref{localCDS} to bound the probability that a typical vertex $V_{i}$ is
pruned by Rule 2. In this case the
blue vertices correspond to nodes in $D_{1}(V_{i})$ with IDs {\em
higher} than $i$, and the white vertices correspond to nodes in
$D_{1}(V_{i})$ with lower IDs. Let $r={1\over \log^{3/2}
\xi_{n}},$ and let ${\cal A}_{i}$ be the event that
$D_{r}(V_{i})\subseteq {\cal Q}_{n}.$ Then
\begin{equation}
\label{calAi} \Pr({\cal A}_{i})={(\ell_{n}-2r)^{2}\over
\ell_{n}^{2}}\geq 1-{4r\over \ell_{n}}. \commentout{1-{4\over
\ell_{n}(\log \xi_{n})^{3/2}}.}
\end{equation}
Let $\hat{D}_{1}(V_{i})=D_{1}(V_{i})\cap {\cal
Q}_{n}$ be the set of points in ${\cal Q}_{n}$ whose distance from
$V_{i}$ is one or less, and let $|\hat{D}_{1}(V_{i})|$ be the area
of $\hat{D}_{1}(V_i).$  Let $\rho_{i}^{(b)}$ denote the number of nodes in $\hat{D}_{1}(V_{i})$
having a label that is larger than $i$, and let $\rho_{i}^{(w)}$
be the number of nodes in $\hat{D}_{1}(V_{i})$ having a label that is
smaller than $i$. Then, given the location of the $i$'th vertex
$V_{i}$,
 $\rho_{i}^{(b)}$ has a Binomial$(n-i,
{|\hat{D}_{1}(V_{i})|\over \ell_{n}^{2}})$ distribution. Define
$\mu_{b}=\mu_{b}(i)$ to be the expected value of $\rho_{i}^{(b)}$
given the location of the $i$'th point:
\begin{equation}
\label{mub} \mu_{b}= E(\rho_{i}^{(b)}| V_{i}
 )=
{(n-i)|\hat{D}_{1}(V_{i})|\over \ell_{n}^{2}}.
\end{equation}
Similarly
 $\rho_{i}^{(w)}$ has a Binomial$(i-1,
{|\hat{D}_{1}(V_{i})|\over \ell_{n}^{2}})$ distribution, and we
define $\mu_{w}=\mu_{w}(i)$ to be the expected value:
\begin{equation}
\label{muw}\mu_{w}=
E(\rho_{i}^{(w)}|V_{i})={(i-1)|\hat{D}(V_i)|\over \ell_{n}^{2}}.
\end{equation}
If ${\cal A}_i$ occurs, then  by Chebyshev's inequality,

\begin{equation}
\label{concrhob}
 \Pr(|\rho_{i}^{(b)}-\mu_{b}(i)|< {\mu_{b}\over 2}|{\cal A}_i)\geq
1-{16\ell_{n}^{2}\over n-i},
\end{equation}
and  similarly for $\rho_{i}^{(w)}.$

If we let ${\cal D}_{i}$ be the event that {\em both} of the
inequalities $|\rho_{i}^{(b)}-\mu_{b}(i)|< {\mu_{b}\over 2}$ and
$|\rho_{i}^{(w)}-\mu_{w}(i)|< {\mu_{w}\over 2}$ are satisfied,
then
\begin{equation}
\label{DgivenA} \Pr({\cal D}_{i}|{\cal A}_{i})\geq
1-{16\ell_{n}^{2}\over n-i}-{16\ell_{n}^{2}\over i-1}.
\end{equation}
Combining (\ref{DgivenA}) and (\ref{calAi}), we get
\begin{equation}
\label{DandA} \Pr({\cal D}_{i}\cap {\cal A}_{i})\geq
\left(1-{16\ell_{n}^{2}\over n-i}-{16\ell_{n}^{2}\over i-1}\right)
\left(1-{4r\over \ell_n}\right).
 \commentout{(1-{4\over \ell_{n}(\log \xi_{n})^{3/2}}).}
\end{equation}

 Now let $\lambda_{n}=n-\alpha_{n},$ then clearly
\begin{equation}
\label{sumoni}
 E({U})\geq
\sum\limits_{i=\alpha_{n}}^{\lambda_{n}}\Pr(I_{i}=1) \geq
\sum\limits_{i=\alpha_{n}}^{\lambda_{n}}\Pr(I_{i}=1|{\cal
D}_{i}\cap {\cal A}_{i})\Pr({\cal D}_{i}\cap {\cal A}_{i})
\end{equation}
To obtain a lower bound for the right hand side of
inequality (\ref{sumoni}), we prove
\begin{lemma}
\label{termtwo} There is a constant $\tilde{c}>0$ such that
for all sufficiently large $n$ and all $\alpha_n\leq i<\lambda_n$,
 $\Pr(I_{i}=1| {\cal D}_{i}\cap {\cal A}_{i})\geq
1-{\tilde{c}\over (\log \xi_{n})^{3/2}}.$
\end{lemma}
\begin{proof}We begin by noting that {\it given} the event
${\cal D}_i\cap{\cal A}_i$ and $\alpha_n\leq i<\lambda_n=
n-\alpha_n$, we have
\begin{equation}
\label{35}
\rho_{i}^{(w)}< {3\over 2}\mu_{w}(i)
={3(i-1)|\hat{D}_1(V_i)|\over 2\ell_n^2}\leq
{3\pi n\over 2\ell_{n}^{2}}.
\end{equation}
Similarly
\begin{equation}
\label{**}
\rho_{i}^{(b)}> {1\over 2}\mu_{b}(i)={(n-i)|\hat{D}_1(V_i)|
\over 2\ell_n^2}> { \alpha_n\pi\over
8\ell_{n}^{2}}={\xi_{n}\pi\over 8}
\end{equation}
It follows from inequalities (\ref{35}) and (\ref{**})
and from the conditions on the sequences
$\langle\xi_n\rangle$ and $\langle\alpha_n\rangle$ that, {\it given} ${\cal D}_i\cap
{\cal A}_i$ and $\alpha_n\leq i<\lambda_n$,
\begin{equation}
  \rho_{i}^{(b)}(\log
\rho_{i}^{(b)})^{3/2}\geq \rho_{i}^{(w)}.
\end{equation}
Next we consider the conditional probability
$\Pr(I_i=1 | \rho_i^{(b)}, \rho_i^{(w)},V_i, {\cal D}_i\cap {\cal A}_i)$
where the values of $\rho_i^{(b)}$ and $\rho_i^{(w)}$ and
the location of  $V_i$ are consistent with the event
${\cal D}_i\cap {\cal A}_i$. In this case, it follows from
inequality (\ref{**}) that
\begin{equation}
\label{***}
\delta(\rho_i^{(b)}) :=
{1\over (\rho_i^{(b)})^{1/3}\log
(\rho_i^{(b)})}\leq {1\over
(\xi_n/3)^{1/3}\log (\xi_n/3)}
\leq {1\over (\log (\xi_n))^{3/2}}=r.
\end{equation}
Since the event ${\cal A}_i$ implies $D_r(V_i)\subseteq\Re_+^2$,
it follows from (\ref{***}) that
$D_{\delta(\rho_i^{(b)})}(V_i)\subseteq\Re_+^2$. Finally,  it
follows from Theorem 4 that for some fixed positive constant
$\tilde{c}$
\begin{equation}
\label{!}
\Pr(I_i=1 | \rho_i^{(b)}, \rho_i^{(w)},V_i, {\cal D}_i\cap {\cal A}_i)
\geq 1-{c\over (\log(\rho_i^{(b)}))^{3/2}}
\geq 1 -{\tilde{c}\over (\log(\xi_i^{(b)}))^{3/2}}
\end{equation}
for all sufficiently large $n$ and all $\alpha_n\leq i<\lambda_n$.
The lemma now follows from (\ref{!}).
\end {proof}

Recall that $\lambda_{n}=n-\alpha_{n}$, that $\alpha_{n}=o(n),$
that $\xi_{n}={\alpha_{n}\over \ell_{n}^{2}}\rightarrow\infty$ as
$n\rightarrow\infty,$ and that for all sufficiently large $n$,
$\alpha_{n}> {16n\over (\log\xi_n)^{3/2}}.$ \commentout{ ${4n\over
\log^{3/2}\xi_{n}} <\alpha_{n}.$ } So it follows from
Lemma \ref{termtwo}
and  (\ref{DandA}) and (\ref{sumoni}), that

$$
E(U)\geq n-2\alpha_{n}+o(\alpha_{n}). $$
\end{proof}

\section{Discussion}

In this final section, assume $\ell_{n}=\Theta(({n\over \log
n})^{t})$ for some fixed positive $t\leq {1\over 2}$.
 For all sufficiently large $n$, the
expected size of the Rule 2 dominating set is at least
$\ell_{n}^{2}/4$ (See Theorem 5 of \cite{HS}).
 There is a gap between this lower
bound and the $O(\alpha_{n})$ upper bound in Theorem \ref{main}.
For example, when $t=1/2$,  the lower and upper bounds for the
expected size of the Rule 2 dominating set are respectively
$\Theta(n/\log n)$ and $\Theta(n/(\log\log n)^{3/2})$.  For
$t<{1/2}$ the gap is even wider: the lower and upper bounds are
respectively $\Theta( ({n\over \log n})^{2t})$ and
$\Theta({n\over \log n}).$ We conjecture that, in fact, the
expected size of the Rule 2 dominating set is
$\Theta(\ell_{n}^{2}).$

\vfill\eject
 \noindent
 {\bf Acknowledgement} We thank Jie Wu for introducing us
to this problem. Professor Wu made helpful suggestions regarding
an early version of this paper.

\end{document}